\documentclass[aps,prd,preprint,superscriptaddress]{revtex4}
\begin{document}
\title{\boldmath Strangeness spin, magnetic moment and
strangeness configurations of the proton}

\author{C. S. An}
\email[]{ancs@ihep.ac.cn} \affiliation{Institute of High Energy
Physics, CAS, P.O.Box 918, Beijing 100049, China}

\author{D. O. Riska}
\email[]{riska@pcu.helsinki.fi} \affiliation{Helsinki Institute of
Physics and Department of Physical Sciences, POB 64, 00014
University of Helsinki, Finland}

\author{B. S. Zou}
\email[]{zoubs@ihep.ac.cn} \affiliation{CCAST (World Lab.),
P.O.Box 8730, Beijing 100080, China} \affiliation{Institute of
High Energy Physics, CAS, P.O.Box 918, Beijing 100049, China}

\date{\today}

\begin{abstract}
The implications of the empirical signatures for the positivity of
the strangeness magnetic moment $\mu_s$ and the negativity of the
the strangeness contribution to the proton spin $\Delta_s$ on the
possible $uuds\bar s$ configurations of five quarks in the proton
are analyzed. The empirical signs for the values for these two
observables can only be obtained in configurations where the
$uuds$ subsystem is orbitally excited and the $\bar s$ antiquark
is in the ground state. The configurations in which the $\bar s$
is orbitally excited, which include the conventional
$K^+\Lambda^0$ configuration, with exception of that in which the
the $uuds$ component has spin 2, yield negative values for
$\mu_s$. Here the strangeness spin $\Delta_s$, the strangeness
magnetic moment $\mu_s$ and axial coupling constant $G_A^s$ are
calculated for all possible configurations of the $uuds\bar s$
component of the proton. In the configuration with $[4]_{FS}[22]_F
[22]_S$ flavor-spin symmetry, which is likely to have the lowest
energy, $\mu_s$ is positive and $\Delta_s \simeq G_A^s \simeq -
1/3 \mu_s$.

\end{abstract}

\maketitle

\section{Introduction}

Four recent experiments on parity violation in electron-proton
scattering suggest that the strangeness magnetic moment of the
proton $\mu_s$ is positive \cite{sample,happex,a4,G0}. In contrast
most theoretical calculations have led to negative values for this
observable \cite{beck}, with but few exceptions
\cite{weigel,isgur,rho,lubov,silva,lewis}. The implications of the
empirical result for the configuration of the constituents of the
proton is considered by a calculation of $\mu_s$ for all positive
parity configurations of the $uuds\bar s$ system with one
constituent in the first orbitally excited state, which may be
contained in the proton, have recently been studied in
Ref.\cite{zou}. The formalism summarized in that reference is
elaborated here in more detail along with a few minor
corrections. In addition it is now applied to the
strangeness contribution to the proton spin $\Delta_s$ and the
corresponding strangeness axial form factor $G_A^s$.

In the case of the strangeness contribution to the proton spin
$\Delta_s$, the empirical indications are that it is very small
\cite{airapet} or small and negative ($-0.10\pm 0.06$)
\cite{Brad,spin}. Extrapolation of the empirical values for the
strangeness axial form factor $G_A^s$ to low $Q^ 2$ indicates a
non-zero negative value for that quantity \cite{Pate}.

Here it is noted that  $\mu_s$ is positive and that both
$\Delta_s$ and $G_A^s$ are negative and smaller than $\mu_s$ in
magnitude in the $uuds\bar s$ configuration, which is likely to
have the lowest energy, where the $\bar s$ quark is in the ground
state and the $uuds$ system is in the $P-$state. If in contrast
the strange antiquark is in the $P-$state and the 4 quarks are in
their ground state the $\mu_s$ has the opposite -- and empirically
contraindicated sign. The latter configurations include that of a
fluctuation of the proton into a kaon and a strange hyperon, which
is well known to lead to a negative value for the strangeness
magnetic moment \cite{musolf,forkel,hannelius,chenxs}.

In this manuscript a detailed description of the wave functions
of the $uuds\bar s$ configurations that may contained in the
proton. The manuscript falls into 6 sections. Section 2 contains the
definitions of the strangeness observables and flavor wave
functions for the $uuds\bar s$ components of the proton along with
a discussion of their expected hyperfine splittings.
The configurations, in which the $uuds$ system is in
the ground state, are considered in section 3 and
the corresponding
configurations, in which the $uuds$ system is in the $P-$state
are considered in section 4.
The implications of the empirical values for $\mu_s$ and
$\Delta_s$ for the probability of the $s\bar s$ in the
proton are considered in section 4. A
summarizing discussion is given in section 5.

\section{The strangeness observables and flavor wave functions
for $uuds\bar s$
components}

In the non-relativistic quark model, the strangeness
magnetic moment $\mu_s$ and the
strangeness contribution to the proton
spin $\Delta_s$ are defined as the expectation value of the
following two operators in the proton state:
\begin{eqnarray}
&&\vec{\mu}_{s}=e\sum_i\frac{\hat{S_{i}}}{2m_{s}}
(\hat{l}_{i}+\hat{\sigma}_{i})\, ,\nonumber\\
&&\vec{\sigma}_{s}=\hat{\sigma}_{s}+
\hat{\sigma}_{\bar{s}}\, ,
\end{eqnarray}
where $\hat{S}_{i}$ is the strangeness counting operator with
eigenvalue $+1$ for s and $-1$ for $\bar{s}$ quark and $m_{s}$ is the
constituent mass of the strange quark.
The strangeness axial form factor is in turn defined as the
matrix element of the strangeness axial current operator
\begin{equation}
\vec A^s = \vec \gamma_i^s \gamma_5^s +
 \vec \gamma_i^{\bar s} \gamma_5^{\bar s} \,.
\end{equation}

The matrix element of this operator in the proton is denoted as
$G_A^ s$. In the non-relativistic limit and at $Q^2 = 0$ one has
$G_A^s = \Delta_s$.

A key ingredient for the calculation of the matrix elements of these
operators are the flavor wave functions for the
$uuds\bar s$ components in the proton. These are usually
constructed by coupling $uuds$ flavor wave functions with
the $\bar s$
flavor wave function. There are four possible flavor symmetry
patterns for the $uuds$ system : $[4]_{F}$, $[31]_{F}$, $[22]_{F}$
and $[211]_{F}$ in the Weyl tableaux of the $SU(3)$ group
\cite{chen,ma}. Combination of these with the anti-quark with flavor
symmetry $[1]^{*}_{F}$ leads to the following pentaquark multiplet
representations of $SU(3)$ :
\begin{equation}
[4]_{F}\otimes[1]^{*}_{F}=\mathbf{10}\oplus\mathbf{35}\, ,
\end{equation}
\begin{equation}
[31]_{F}\otimes[1]^{*}_{F}=\mathbf{8}\oplus\mathbf{10}
\oplus\mathbf{27}\, ,
\end{equation}
\begin{equation}
[22]_{F}\otimes[1]^{*}_{F}=\mathbf{8}\oplus\mathbf{\bar{10}}\, ,
\end{equation}
\begin{equation}
[211]_{F}\otimes[1]^{*}_{F}=\mathbf{1}\oplus\mathbf{8}\, .
\label{n4}
\end{equation}
Here the numbers in boldface on the right-hand-side of
the equations indicate the
dimensions of the pentaquark representations.
As an example the tentative $\theta^+$-pentaquark belongs to the
baryon anti-decuplet $\mathbf{\bar{10}}$ representation. Since the
proton belongs to the baryon octet representation, the possible
flavor symmetry patterns for the $uuds$ in the proton are limited
to $[22]_{F}$, $[211]_{F}$ and $[31]_{F}$. The corresponding
flavor wave functions can be obtained as in refs.
\cite{chen,ma}. For convenience we list the relevant ones here
explicitly.

For the mixed flavor symmetry representations
$[22]_F$ and $[211]_{F}$, there are 2 and
3 independent flavor wave functions, respectively:
\begin{eqnarray}
|[22]_{F_{1}}\rangle \!&=&
\frac{1}{\sqrt{24}}[2|uuds\rangle+2|uusd\rangle+2|dsuu\rangle+2|sduu\rangle
-|duus\rangle-|udus\rangle\nonumber\\
&&-|sudu\rangle-
|usdu\rangle-|suud\rangle-|dusu\rangle-|usud\rangle
-|udsu\rangle]\, , \label{combf1}\\
|[22]_{F_{2}}\rangle \!&=&\!
\frac{1}{\sqrt{8}}[|udus\rangle+|sudu\rangle+|dusu\rangle+|usud\rangle-
|duus\rangle-|usdu\rangle\nonumber\\
&&-|suud\rangle-|udsu\rangle], \quad
\label{combf2}\\
|[211]_{F_{1}}\rangle \!&=&
\frac{1}{\sqrt{16}}[2|uuds\rangle-2|uusd\rangle
-|duus\rangle-|udus\rangle-|sudu\rangle-|usdu\rangle\nonumber\\
&&+|suud\rangle+|dusu\rangle+|usud\rangle
+|udsu\rangle]\, , \label{combf3}\\
|[211]_{F_{2}}\rangle \!&=&
\frac{1}{\sqrt{48}}[3|udus\rangle-3|duus\rangle
+3|suud\rangle-3|usud\rangle+2|dsuu\rangle-2|sduu\rangle\nonumber\\
&&-|sudu\rangle+|usdu\rangle+|dusu\rangle-|udsu\rangle]\, , \label{combf4}\\
|[211]_{F_{3}}\rangle \!&=&
\frac{1}{\sqrt{6}}[|sudu\rangle+|udsu\rangle
+|dsuu\rangle-|usdu\rangle-|dusu\rangle-|sduu\rangle]\, .
\label{combf5}
\end{eqnarray}

In the case of the mixed flavor symmetry $[31]_{F}$ for the $uuds$
system there is a need to separate the isospin 1/2 and 3/2 states,
both of which are listed in Ref~\cite{chen}. The reason for the
presence of two separate classes of wave functions with the flavor
$[31]_{F}$ is the following. In the case of flavor $SU(3)$ model,
the $\Delta^{+}$ resonance is composed $uud$ valence quarks with
the flavor symmetry $[3]_{F}$. In the consideration of the
$uuds\bar{s}$ component of this baryon, and its $uuds$ subsystem,
the latter may also have the mixed flavor symmetry $[31]_{F}$.
This flavor state of the $uuds$ subsystem is therefore a
combination of a $\Delta^+$ and a proton component. There are 6
independent flavor wave functions for $uuds$ of the flavor
symmetry $[31]_{F}$. Among them, by using the weight diagram
method ~\cite{ma}, the three wave functions labeled as
$\psi^{\theta}$ in Ref~\cite{chen} are found to correspond to
isospin 1/2. These are explicitly:
\begin{eqnarray}
|[31]_{F_{1}}\rangle \!&=&
\frac{1}{\sqrt{18}}[2|uusd\rangle+2|suud\rangle
+2|usud\rangle-|sudu\rangle-|usdu\rangle-|dusu\rangle\nonumber\\
&&-|udsu\rangle-|dsuu\rangle-|sduu\rangle]\, , \label{combf6}\\
|[31]_{F_{2}}\rangle \!&=&
\frac{1}{12}[6|uuds\rangle-3|duus\rangle
-3|udus\rangle-4|dsuu\rangle-4|sduu\rangle+5|sudu\rangle\nonumber\\
&&+5|usdu\rangle+2|uusd\rangle-|suud\rangle-|dusu\rangle-|usud\rangle
-|udsu\rangle]\, , \label{combf7}\\
|[31]_{F_{3}}\rangle \!&=& \frac{1}{\sqrt{48}} [-3|duus\rangle
+3|udus\rangle-3|dusu\rangle
+3|udsu\rangle-2|dsuu\rangle+2|sduu\rangle\nonumber\\
&&-|sudu\rangle+|usdu\rangle -|suud\rangle+|usud\rangle]\ .
\label{combf8}
\end{eqnarray}

The color symmetry of all the $uuds$ configurations is limited to
be $[211]_C$ in order to combine with the $\bar s$ antiquark to
form a color singlet (cf.(\ref{n4})). The Pauli principle requires
that corresponding orbital-flavor-spin states have the mixed
symmetry $[31]_{XFS}$ so as to combine with the mixed color
symmetry state $[211]_C$ to form the required completely
antisymmetric 4-quark state $[1111]$. Since the intrinsic parity
is positive for a quark and negative for an anti-quark, the
$uuds\bar s$ component in a proton, which has positive parity,
requires that either the $\bar s$ is in the P-state with the
$uuds$ system in the ground state with spatial symmetry $[4]_X$ or
that one of the quarks is in the P-state so that
the $uuds$ system has mixed spatial
symmetry $[31]_X$.

There are several configurations of the $uuds\bar s$ system that
have positive parity, isospin 1/2, spin 1/2 and one unit of
orbital angular momentum. The spin-dependent hyperfine interaction
between the quarks splits these states, so that the configurations
with the lowest energy may be expected to be those with highest
probability for admixture in the proton.

In most models for the hyperfine interaction between quarks
in the baryon is spin dependent. In the common colormagnetic
 hyperfine
interaction model, the spin dependence is such that the spin
singlet state has lower energy than the spin triplet state
\cite{capstick}.
This is also the case for the instanton induced
interaction model \cite{metsch}. Finally, the schematic flavor and
spin dependent interaction model:
\begin{equation}
H' = -C\sum_{i<j}\vec\lambda_i\cdot\vec\lambda_j\,
\vec\sigma_i\cdot\vec\sigma_j\, ,
\label{flavorspin}
\end{equation}
which gives the qualitative description with correct ordering of
the states in low lying part of the baryon spectrum in all flavor
sectors \cite{gloz1,gloz2}, implies that antisymmetric flavor and
spin configurations have the lowest energy. In this interaction
$C$ is a constant that represents an average of the spin-spin part
of the interaction expected to be mediated by pseudoscalar meson
exchange \cite{brown}. Phenomenologically $C \sim 20 - 30$ MeV.

In next two sections, the strangeness spin $\Delta_s$ and magnetic
moment $\mu_s$ are calculated for all possible configurations of
the $uuds\bar s$ system that have the quantum numbers of the proton.

\section{The configurations with the $uuds$ system in its ground state}

In the configurations where the $uuds$ quarks are in their ground
state, the spatial state has complete symmetry $[4]_X$, and
accordingly their
flavor-spin state has to have the mixed symmetry $[31]_{FS}$. All
the different flavor and spin state symmetry configurations that
can combine to the required $[31]_{FS}$ mixed symmetry combination
have been listed in Table \ref{table1} \cite{helminen}. In the
table the matrix elements of the quadratic Casimir
operator of $SU(6)$, $C_2^{(6)}$, multiplied by the constant $C$ in the
flavor-spin interaction (\ref{flavorspin}) are also listed so as
to give an indication of the energy splitting between these
configurations. The requirement of positive parity requires that
for these configurations the strange antiquark in the $uuds\bar s$
system has to be in the $P-$state.

\begin{table}[hb]\caption{\footnotesize Flavor
and spin configurations of the \emph{ uuds} quark states in the
ground state~\cite{helminen} and the corresponding operator matrix
elements .}\vspace{0.5cm}\footnotesize
\begin{tabular} {l|c|c|c|c}
\hline
$uuds$
ground
state & $\bar{\sigma}_{s}$ & $\Delta_{s}(P_{s\bar{s}})$ & $\mu_{s}
(\frac{m_{p}}{m_{s}}P_{s\bar{s}})$&$-C\,C_2^{(6)}$\\
\hline
$[31]_{FS}[211]_{F}[22]_{S}$& -- &$-1/3$ & $-1/3$&$-16C$\\
$[31]_{FS}[211]_{F}[31]_{S}$&$13/36$&$85/216$&$-95/216$&$-13
{1\over 3}C$\\
$[31]_{FS}[22]_{F}[31]_{S}$&$1/2$&$5/12$&$-5/12$&$-9{1\over 3}C$\\
$[31]_{FS}[31]_{F}[22]_{S}$&--&$-1/3$&$-1/3$&$-8C$\\
$[31]_{FS}[31]_{F}[31]_{S}$&$65/108$&$281/648$&$-259/648$&$-5{1\over 3}C$\\
$[31]_{FS}[31]_{F}[4]_{S}$&--&$1/6$&$7/6$&$0$ \\ \hline
\end{tabular}
\label{table1}
\end{table}

The wave functions of the proton state with spin +1/2 that have
any one of these symmetries may be written in the general form:
\begin{equation}
|p,+{1\over 2}\rangle=
A_{s\bar{s}}\sum_{abc}\sum_{s_{z}mm's'_{z}}C^{\frac{1}{2}
\frac{1}{2}}_{1s_{z},jm}C^{jm}_{1m',\frac{1}{2}s'_{z}}C^{[1^4]}_{[31]_{a}
[211]_{a}}C^{[31]_{a}}_{[F]_{b}[S]_{c}}[F]_{b}[S]_{c}[211]_{C,a}
\bar Y_{1m'} \bar \chi_{s_{z'}}\varphi(\{\vec r_i\})\, .
\label{wfc1}
\end{equation}
Here $[F]_{b},[S]_{c}$ and $[211]_{C,a}$ represent the flavor,
spin and color state wave functions, denoted by their Weyl
tableaux respectively. The sums over $a$, $b$ and $c$ run over the
configurations of the $[31][F][S]$ representations of $S_{4}$
permutation group for which the corresponding Clebsch-Gordan
coefficients $C^{a}_{b,c}$ do not vanish. The value of the first
of these CG coefficients are $C^{[1^4]}_{[31]_{a}
[211]_{a}}=\pm\frac{1}{\sqrt{3}}$~\cite{chen}. Finally  $\bar
Y_{1m'}$ and $\bar \chi_{s'_{z}}$ denote the orbital and spin
states of the anti-strange quark respectively. In Eq.(\ref{wfc1})
$A_{s\bar{s}}$ is the amplitude of the $s\bar{s}$ component in the
proton and $\varphi(\{\vec r_{i}\})$ is a symmetric function of
the coordinates of the $uuds\bar{s}$.

The wave functions of the $uuds$ subsystem may be organized in
groups according to their spin states. Consider first the states
which have the spin symmetry $[22]_S$. To these belong the
configuration $[211]_F[22]_S$, which is expected to have the
lowest energy among all the configurations with the mixed
flavor-spin symmetry $[31]_{FS}$. As the total spin of the $uuds$
system with this symmetry is $S=0$, and the angular momentum space
is isotropic, it gives no contribution to $\mu_s$ and
$\sigma_{s}$. In these configurations only the $\bar{s}$ quark
contributes to $\mu_s$ and $\Delta_s$ (and $G_A^ s$):
\begin{eqnarray}
\mu_{s}&=&-\frac{1}{3}\frac{e}{2m_{s}}P_{s\bar{s}}\, ,
\\ \Delta_{s}&=&-\frac{1}{3}P_{s\bar{s}}\, .
\end{eqnarray}
Here $P_{s\bar s}$ is the probability of this configuration.
In units of the nuclear magnetons $\mu_s$ takes the value:
\begin{equation}\mu_{s}=-\frac{1}{3}\frac{m_{p}}{m_{s}}
P_{s\bar{s}}\, .
\end{equation}
In this configuration the numerical value of $\mu_s$
is negative and $\sim$ 2 times $\Delta_s$, as $m_p \simeq 2 m_s$.

The spin of the $uuds$ system in the states which have the spin
symmetry $[31]_{S}$ is $S=1$. If we neglect the interaction of
quarks, then the wave function of the proton in the angular
momentum space may be written in the general form:
\begin{eqnarray}&& |p,+{1\over 2}\rangle
=\frac{A_{s\bar{s}}}{\sqrt{2}}\{
C_{11,\frac{1}{2}-\frac{1}{2}}^{\frac{1}{2}\frac{1}{2}}
[C_{11,10}^{11}|11 \rangle_{A_{S}}|10\rangle_{B_{X}}
+C_{10,11}^{11}|10\rangle_{A_{S}}|11\rangle_{B_{X}}]|\frac{1}{2}
\frac{1}{2}\rangle_{B_{S}}  \nonumber\\
&& + C_{10,\frac{1}{2}\frac{1}{2}}^{\frac{1}{2}\frac{1}{2}}
[C_{11,1-1}^{10}|11\rangle_{A_{S}}|1\!-\!\!1 \rangle_{B_{X}}
+C_{1-1,11}^{10}|1\!-\!\!1\rangle_{A_{S}}|11\rangle_{B_{X}}]|\frac{1}{2}
\frac{1}{2}\rangle_{B_{S}}+ 
\label{wfc2}\\
&&C_{00,\frac{1}{2}
\frac{1}{2}}^{\frac{1}{2}\frac{1}{2}}  [C_{11,1-1}^{00}|11\rangle
_{A_{S}}|1\!-\!\!1\rangle_{B_{X}}+C_{10,10}^{00}|10\rangle_{A_{S}}|10
\rangle_{B_{X}}+
C_{1-1,11}^{00}|1\!-\!\!1\rangle_{A_{S}}|11\rangle_{B_{X}}]|\frac{1}{2}
\frac{1}{2}\rangle_{B_S} \} \nonumber\, .
\end{eqnarray}
Here for the sake of abbreviation the subindica $A_S$ and $B_X$ represent
the spin state of the $uuds$ system and the
orbital state of the $\bar{s}$ quark respectively. Explicit evaluation
of the matrix elements with these wave functions for $\mu_s$ and
$\Delta_s$ leads to the results:
\begin{eqnarray}
&&\mu_{s} =
-{1\over 2} \frac{m_p}{m_s}
(1-\frac{1}{3}\bar\sigma_s  )P_{s\bar s}\, .
\label{firstA} \\
&& \Delta_s  =  {1\over 3}
(1+\frac{1}{2}\bar \sigma_s) P_{s\bar s}\, .
\label{avspin}
\end{eqnarray}
Here the $\bar\sigma_{s}$ is the expectation value of the
z-component of the spin of the $s$ quark in the configuration
where $s_{z}=1$. (Note that Eq.(\ref{firstA}) here corrects Eq.(4)
in \cite{zou}). The numerical value of $\bar\sigma_s$ depends on
the detailed configuration as shown below. The results in all
cases satisfy the inequality $\bar\sigma_s < 1$ as shown in Table
\ref{table1}. This result implies that $\mu_s$ in all these
configuration is negative and that $\Delta_s$ is positive, in
contradiction with experiment.

The final possibility when the spatial state of the $uuds$ system
is that the spin state of the $uuds$ system is completely
symmetric: $[4]_{S}$. In this case it has spin $S=2$ and as a
consequence the total angular momentum of the $\bar{s}$ quark has
to be $j=\frac{3}{2}$ in order to combine with the $uuds$ system
to form the proton state with spin +1/2. The wave function then
may be expressed in a way analogous to Eq.(\ref{wfc2}). The
relevant matrix elements for these configurations are found to be
:
\begin{eqnarray}
\mu_{s}&=&\frac{7}{6}\frac{m_{p}}{m_{s}}P_{s\bar{}s}\, ,\\
\Delta_{s}&=&\frac{1}{6}P_{s\bar{s}}\, .
\end{eqnarray}
This configuration thus yields positive values for both $\mu_s$
and $\Delta_s$ as well as $G_A^s$. This configuration is very
unlikely to have a large probability of the proton, as it is
expected to have an energy of 100 - 150 MeV above all the other
configurations with the mixed flavor-spin symmetry $[31]_{FS}$.

The (tedious) calculation of the average spin value $\bar\sigma_s$
in Eq.(\ref{avspin}) may illustrated by the following explicit
calculation for the case of the configuration
$[31]_{FS}[22]_F[31]_S$. The results for this and the other
configurations are listed in Table \ref{table1}.

There are three combinations of the mixed symmetry states $[22]_F$
and $[31]_S$ that combine to the mixed symmetry state $[31]_{FS}$.
The explicit expressions for these are the following \cite{chen}:
\begin{eqnarray}|[31]_{FS_{1}}\rangle&=&\frac{1}{\sqrt{2}}\{|[22]_{F_{1}}
\rangle|[31]_{S_{2}}\rangle+|[22]_{F_{2}}|[31]_{S_{3}}\rangle\}\,
 ,\label{comb1}\\
|[31]_{FS_{2}}\rangle&=&\frac{1}{2}\{\sqrt{2}|[22]_{F_{1}}
|[31]_{S_{1}}\rangle
+|[22]_{F_{1}}\rangle|[31]_{S_{2}}\rangle-|[22]_{F_{2}}
\rangle|[31]_{S_{3}}\rangle\}\, ,\label{comb2}\\
|[31]_{FS_{3}}\rangle&=&\frac{1}{2}\{\sqrt{2}|[22]_{F_{2}
}\rangle|[31]_{S_{1}}\rangle-
|[22]_{F_{2}}\rangle|[31]_{S_{2}}\rangle-|[22]_{F_{1}
}\rangle|[31]_{S_{3}}\rangle\}\, . \label{comb3}
\end{eqnarray}
where the $[22]_F$ flavor functions are given by
Eqs.(\ref{combf1},\ref{combf2}) and the three spin wave functions
are
\begin{eqnarray}
|[31]_{S_{1}}>&=&\frac{1}{\sqrt{12}}[3|\uparrow\uparrow
\uparrow\downarrow\rangle-|\uparrow\uparrow\downarrow\uparrow\rangle
-|\uparrow\downarrow\uparrow\uparrow\rangle-|\downarrow
\uparrow\uparrow\uparrow\rangle]\, ,\label{combs1}\\
|[31]_{S_{2}}>&=&\frac{1}{\sqrt{6}}[2|\uparrow\uparrow
\downarrow\uparrow\rangle-|\uparrow\downarrow\uparrow
\uparrow\rangle-|\downarrow\uparrow\uparrow\uparrow\rangle]\, ,\label{combs2}\\
|[31]_{S_{3}}>&=&\frac{1}{\sqrt{2}}[|\uparrow\downarrow
\uparrow\uparrow\rangle-|\downarrow\uparrow\uparrow\uparrow\rangle]\,.
\label{combs3}
\end{eqnarray}
From Eqs.(\ref{comb1}-\ref{comb3}) one obtains
\begin{equation}\bar{\sigma}_{s}=\frac{1}{3}
(\bar{\sigma}_{FS_{1}}+\bar{\sigma}_{FS_{2}}+\bar{\sigma}_{FS_{3}})\, .
\end{equation}
It then follows from
Eqs.(\ref{combf1},\ref{combf2},\ref{combs1}-\ref{combs3}) that
\begin{equation}\bar{\sigma}_{s}=\frac{1}{2}\, ,\end{equation}
and at last, from Eqs.(21,22),
\begin{eqnarray}
\mu_{s}&=&-\frac{5}{12}\frac{m_{p}}{m_{s}}P_{s\bar{s}}\, ,\\
\Delta_{s}&=&\frac{5}{12}P_{s\bar{s}}\, .
\end{eqnarray}

The matrix elements needed for the values of $\Delta_{s}$ and
$\mu_{s}$ in all other configurations may be calculated by the
same method.

\section{The configurations with the $\bar s$ in its ground
state}

In the configurations, where the $\bar s$ antiquark is in the
ground state, the lowest possible orbital configuration allowed by
the requirement of positive parity for the $uuds$ state is that
with orbital angular momentum of $L=1$. The corresponding spatial
state has the mixed symmetry labeled by the Weyl tableau
$[31]_{X}$. The possible symmetries of the flavor-spin state are
then either complete symmetry, $[4]_{FS}$, or the mixed
symmetries $[31]_{FS}$, $[22]_{FS}$ and $[211]_{FS}$. All the
flavor and spin symmetry configurations that can combine to these
configurations, and which can be a component of the proton, are
listed in Table II. The wave functions of the proton with spin
+1/2 may for all of these symmetry configurations be expressed in
the general form:

\begin{equation}\begin{array}{ll}|p, +{1\over 2}\rangle=
&\sum_{abcde}\sum_{Ms'_{z}ms_{z}}C^{\frac{1}{2}\frac{1}{2}}
_{JM,\frac{1}{2}s'_{z}}C^{JM}_{1m,Ss_{z}}C^{[1^{4}]}_{[31]_{a}
[211]_{a}}C^{[31]_{a}}_{[31]_{b}[FS]_{c}}
C^{[FS]_{c}}_{[F]_{d}[S]_{e}}\\
&[31]_{X,m}(b)
[F]_{d}[S]_{s_{z}}(e)[211]_{C}(a)\bar\chi_{s'_{z}}
\varphi(\{r_i\})
\, .
\label{wfc3}
\end{array}\end{equation}
Here J is the total angular momentum of the $uuds$ system
M the corresponding
z-component. The orbital angular momentum of these $uuds$ states
is
$L=1$, with the z-component $m$.

\begin{table}[hb]\begin{caption}{\footnotesize{Flavor
and spin configurations of the\emph{ uuds} quark states in the
first orbitally excited state~\cite{helminen} with total angular
momentum $J=1$ and the corresponding operator matrix elements.
}}\end{caption}\vspace{0.5cm}\footnotesize
\begin{tabular}[b]{l|c|c|c|c|c}
\hline
$uuds$ p-state&$\bar{l}_{s}$&
$\bar{\sigma}_{s}$&$\Delta_{s}(P_{s\bar{s}})$&$\mu_{s}(\frac{m_{p}}{m_{s}}
P_{s\bar{s}})$ & $-C C_2^{(6)}$ \\
\hline
$[4]_{FS}[22]_{F}[22]_{S}$&$1/4$&--&$-1/3$&$1/2$&$-28C$\\
$[4]_{FS}[31]_{F}[31]_{S}$&$1/4$&$7/9$&$-2/27$&$73/108$&$-21{1\over 3}C$\\
$[31]_{FS}[211]_{F}[22]_{S}$&$13/48$&--&$-1/3$&$37/72$
&$-16 C$\\
$[31]_{FS}[211]_{F}[31]_{S}$&$119/432$&$13/36$&$-23/108$&$707/1296$
&$-13{1\over 3}C$\\
$[31]_{FS}[22]_{F}[31]_{S}$&$1/4$&$1/2$&$-1/6$&$7/12$
&$-7{1\over 3}C$\\
$[31]_{FS}[31]_{F}[22]_{S}$&$13/48$&--&$-1/3$&$37/72$
&$-8C$\\
$[22]_{FS}[211]_{F}[31]_{S}$&$1/4$&$5/12$&$-7/36$&$5/9$
&$-5{1\over 3}$\\
$[31]_{FS}[31]_{F}[31]_{S}$&$295/1296$&$65/108$&$-43/324$
&$2371/3888$&$-5{1\over 3}$\\
$[22]_{FS}[22]_{F}[22]_{S}$&$1/4$&--&$-1/3$&$1/2$
&$-4 C$\\
$[211]_{FS}[211]_{F}[22]_{S}$&$11/48$&--&$-1/3$&$35/72$
&$0$\\
$[31]_{FS}[31]_{F}[4]_{S}$&$43/216$&--&$1/6$&$497/648$
&$0$\\
$[211]_{FS}[211]_{F}[31]_{S}$&$119/432$&$65/108$&$-43/324$&$811/1296$
&$2{1\over 3}C$\\
$[22]_{FS}[31]_{F}[31]_{S}$&$13/54$&$5/12$&$-7/36$&$251/324$
&$2{2\over 3}C$\\
$[22]_{FS}[22]_{F}[4]_{S}$&$1/4$&--&$1/6$&$3/4$
&$4C$\\
$[211]_{FS}[22]_{F}[31]_{S}$&$1/4$&$1/2$&$-1/6$&$7/12$
&$6{2\over 3}$\\
$[211]_{FS}[211]_{F}[4]_{S}$&$23/72$&--&$1/6$&$157/216$
& $8C $ \\
$[211]_{FS}[31]_{F}[22]_{S}$&$11/48$&--&$-1/3$&$35/72$&$8C$\\

$[211]_{FS}[31]_{F}[31]_{S}$&$31/144$&$13/36$&$-23/108$&$227/432$
&$10{2\over 3}C$\\\hline
\end{tabular}
\label{table2x}
\end{table}

These configurations may also be organized in groups
according to their spin
symmetry. The method described explicitly in the previous section
may then be applied to evaluate the matrix elements that
are required for $\mu_s$ and $\Delta_s$.

Of these configurations, the configuration $[4]_{FS}[22]_F[22]_S$
is expected to have the lowest energy. For this and all the
configurations that have the mixed spin symmetry $[22]_{S}$
the desired matrix
elements for $\mu_s$ and $\Delta_s$ are:
\begin{eqnarray}
\mu_s &=& \frac{m_{p}} {3 m_{s}}
(1+ 2 \bar{l}_{s}) P_{s\bar{s}}
\, ,\\
\Delta_s &=& -\frac{1}{3} P_{s\bar{s}}\, .
\label{lowa}
\end{eqnarray}
Here the $\bar{l}_{s}$ is the average value of the z-component of
the orbital angular momentum of the $s$ quark in the $uuds$ system
for $m=1$. The numerical value for $\bar l_s$ depends on the
detailed configuration and may be calculated by the same method as
that described above for the calculation of $\bar \sigma_s$ in the
previous section. The results of $\bar{l}_{s}$ for various
$[31]_X$ configurations with $uuds$ total angular momentum $J=1$
are listed in Table II along with the corresponding values of
$\bar \sigma_s$. The values for $\bar l_s$ are in every case
smaller than 1/2: $\bar l_s < 1/2$. Note that the values
of $\bar{l}_{s}$ and $\bar\sigma_s$ given in Ref.\cite{zou}
for the mixed $[31]_F$
flavor symmetry are not correct due to the neglect of the isospin
decomposition and are corrected in Table II.

In the case of the lowest energy configuration $[4]_{FS}[22]_F[22]_S$
$\bar l_s=1/4$ and consequently
\begin{equation}
\mu_s = \frac{m_{p}} {2 m_{s}} P_{s\bar{s}}\,.
\label{lowest}
\end{equation}
The value of $\mu_s$ in this configuration is consequently
positive, and since $m_p/m_s\simeq 2$, it is approximately
equal to the probability of that configuration in the proton.
Moreover for this configuration $\mu_s \simeq -3\Delta_s$,
a relation that agrees with the present experimental
values within their uncertainty range.

Because of its low energy, this is the most likely $s\bar s$
component in the proton. The $[4]_{FS}[22]_F[22]_S$ $uuds$ wave
function has the simple form:
\begin{equation}
|[4]_{FS}\rangle = {1\over \sqrt{2}}\{[22]_{F_1}[22]_{S_1}
+[22]_{F_2}[22]_{S_2}\}\, .
\label{lowesta}
\end{equation}
The explicit form for the two flavor components $[22]_{F1}$ and
$[22]_{F_2}$ are given in Eqs.(\ref{combf1},\ref{combf2}). The two
corresponding spin functions are readily constructed by the
substitutions $u\leftrightarrow \uparrow$ and $d,s\leftrightarrow
\downarrow$ with an additional $1/\sqrt{2}$ in the normalization
factor. To complete the wave function for this configuration one
needs the explicit antisymmetric color-space wave function:
\begin{equation}
|[1111]_{CX}[211]_C[31]_X ;m\rangle = \sum_{a=1}^3
C^{[1^4]}_{[31]_{a} [211]_{a}}[211]^a_C [31]^a_{Xm}\, .
\label{lowestb}
\end{equation}
As the operators that are considered here do not depend on color,
the explicit color wave functions with the mixed symmetry
$[211]_C$ are not required. The explicit spatial wave functions
may be constructed by reference to
Eqs.(\ref{combs1}-\ref{combs3}), with the substitution of ground
state wave functions for the 3 constituents denoted
 $\uparrow$ and a $P-$state wave function, multiplied
by the spherical harmonic
$Y_{1m}$, for the constituent denoted $\downarrow$.

For the states which have spin symmetry $[31]_{S}$ the total
angular momentum of the $uuds$ system $J$ may take the values 0
and 1. For $J=0$ the results are:
\begin{eqnarray}
\mu_{s}&=&-\frac{m_{p}}{m_{s}}P_{s\bar{s}}\, ,\\
\Delta_{s}&=&P_{s\bar{s}}\, .
\end{eqnarray}
These are the only configurations with the $uuds$ system in an
orbitally excited state, which lead to negative values for the
strangeness magnetic moment. As the lowest one of these
configurations lies $\sim$ 140 - 200 MeV above the configuration
with $[4]_{FS}[22]_F[22]_S$ symmetry, it is unlikely to have a
large probability as a component of the proton.

In the case when $J=1$ the corresponding results are
\begin{eqnarray}
\mu_{s}&=&\frac{m_p}{3 m_s}
(1+\bar{l}_{s}+\bar{\sigma}_{s})P_{s\bar s}\, ,\\
\Delta_{s}&=& -\frac{1}{3}(1-\bar{\sigma}_{s}) P_{s\bar s}\,.
\end{eqnarray}

Finally for the states which have spin symmetry $[4]_{S}$ the results
are:
\begin{eqnarray}
\mu_s  & = &{1\over 3}\frac{m_p}{m_s}
(\frac{5}{2}-\bar l_s)P_{s\bar{s}}\, ,\\
\Delta_s& =& \frac{1}{6}P_{s\bar s}\, .
\end{eqnarray}
As the strangeness contribution to the proton spin is
positive for these configurations in apparent conflict with
the experimental situation, as these are expected to require
a relatively high energy of excitation, they are not
likely to have a large probability in the proton.

\section{The probability of the $uuds\bar s$ component}

Above the possibility for transition matrix elements between
the $uuds\bar s$ and the $uud$ components of the proton
was not considered, as only diagonal matrix elements in
configurations with $s\bar s$ components were calculated.
The calculation of such transition matrix elements demands
an explicit model for the spatial wave functions for
determination of the overlap. These transition matrix elements
modify the proportionality coefficients
between the strangeness magnetic moment and the
probability of the $s\bar s$ components in the proton.
Referring to the analogous situation in the case of decay
of the $\Delta(1232)$ considered in ref.\cite{QBL}, these
contributions are expected to allow a smaller values
for $P_{s\bar s}$ than eg. Eq.(\ref{lowest}) with the present empirical
values for $\mu_s$. As an example in the case of the
$[31]_X[4]_{FS}[22]_F[22]_S$
configuration, the effect of the transition matrix elements
would lead to modification of the expression (\ref{lowest}) for
$\mu_s$ by an additional factor:
\begin{equation}
\mu_s = {m_p\over 2 m_s} (1+\delta)P_{s\bar s}\, .
\label{mod1}
\end{equation}
Here the relative modification caused by transition matrix
elements is contained in the term $\delta$.
Because the $uuds$ configuration is a $P-$state,
this term would have the general form:
\begin{equation}
\delta \sim C_0 m_s \sqrt{<r^2>}\sqrt{P_{uud}\over P_{s\bar s}} \,
, \label{estimate}
\end{equation}
where $P_{uud}+P_{s\bar s}$. Here $C_0$ is a factor that takes
into account the overlap between the wave function for the
$uud\gamma$ state and the $uuds\bar s$ component of the proton,
and $<r^2>$ is the mean square radius of the $uuds\bar s$
component. The mass factor $m_s$ arises from the $s\bar s 
\rightarrow
\gamma$ vertex. Provided that these wave functions have the same phase,
it follows from (\ref{mod1}) that the empirical value for $\mu_s$
sets an upper limit for $P_{s\bar s}$. As an example the
harmonic oscillator model employed in ref.\cite{QBL}, in which
$\sqrt{<r^2>}=1/\omega$, the value for the coefficient
$C_0$ is $C_0=3/8\simeq 0.37$. Since $2 m_s\sim m_p$ and in the
oscillator model $m_s\sqrt{<r^2>}=m_s/\omega\sim 1...1.5$ the following
approximate relation emerges for $\mu_s$ in the
$[31]_X[4]_{FS}[22]_F[22]_S$ configuration:
\begin{equation}
\mu_s\simeq P_{s\bar s} + (0.37...0.55) \sqrt{P_{s\bar s}
-P_{s\bar s}^2}\, . \label{estimx}
\end{equation}
If this equation is solved for the $s\bar s$ configuration
probability $P_{s\bar s}$ with the mean result for $\mu_s$ given
by the SAMPLE experiment ($\mu_s=0.37\pm 0.2\pm 0.26 \pm 0,07$)
\cite{sample} as input one obtains $P_{s\bar s}\simeq
0.17...0.22$. By (\ref{lowa}) this probability for 
the $s\bar s$ configuration gives
$\Delta_s=-0.06...-0.07$, which values falls well within the range
of values for this quantity ($-0.10\pm 0.06$) that is determined
from the recently enlarged set of inclusive and semi-inclusive
polarized deep inelastic scattering data \cite{spin}. Note that in
the case of $\Delta_s$ or $G_A^ s$, the contribution from
transition matrix elements relative to the diagonal matrix
elements is much smaller than in the case of $\mu_s$.

The inclusion of the $[31]_X[4]_{FS}[22]_F[22]_S$ type of
pentaquark components not only reproduces well the strangeness
magnetic moment and spin of the proton, but also 
is consistent with the
observed excess of $\bar d$ over $\bar u$ \cite{Garvey} and the
quark spin contribution \cite{spin} in the proton. In this
configuration, only $udud\bar d$ and $udus\bar s$ pentaquark
components are allowed for the proton but no $uduu\bar u$
component. The quark wave function for the proton may then be
expressed as:
\begin{equation}
|p> = A_{3q}|uud> + A_{d\bar d} |[ud][ud]\bar d> + A_{s\bar s}
|[ud][us]\bar s>\, ,
\end{equation}
with the normalization condition $|A_{3q}|^2+|A_{d\bar d}|^2+
|A^2_{s\bar s}|^2=1$. To reproduce the observed light flavor sea
quark asymmetry in the proton, $\bar d-\bar u=0.12$ \cite{Garvey},
one then needs $P_{d\bar d}\equiv |A_{d\bar d}|^2 =12\%$, while to
reproduce the strangeness spin of the proton, $\Delta_s=-0.10\pm
0.06$ \cite{spin}, one needs $P_{s\bar s}=(12-48)\%$. With more
than $24\%$ of this kind of pentaquark component in the proton,
this can reproduce the observed $u$-quark spin contribution in the
proton $\Delta_u=0.85\pm 0.17$ \cite{spin} as well, since the
pentaquark components give no contribution to the $\Delta_u$ and
$\Delta_u={4\over 3}|A_{3q}|^2$. Instead they give a contribution:
$\Delta L_q={4\over 3}(P_{d\bar d}+P_{s\bar s})$ of the proton
spin through the orbital angular momentum of quarks. In
Ref.\cite{spin}, the spin contributions from both $\bar d$ and
$\bar s$ antiquarks were found to be negative, while that from
$\bar u$ is very uncertain even as to its sign and could be far
less polarized than $\bar d$ and $\bar s$. The pentaquark
configuration considered here gives such quark spin contributions, with
$-{1\over 3}P_{d\bar d}$ from $\bar d$ and $-{1\over 3}P_{s\bar
s}$ from $\bar s$ and zero from $\bar u$ antiquarks. The total d-quark spin
contribution is predicted to be $\Delta_d=-{1\over 3}(1-P_{s\bar
s})$ which is only slightly smaller than the lower end of the
observed value $\Delta_d=-(0.33\sim 0.56)$. In Ref.\cite{spin}, it
was also found that the available data do not require the 
gluon to be polarized.

\section{Discussion}

The complete analysis given above of all the positive parity
configurations of the form $uuds\bar s$ with spin 1/2 and an at
most one unit of orbital angular momentum reveals that the present
empirical signs for $\mu_s$ and $\Delta_s$ imply that the $uuds$
subsystem has to be orbitally excited and that the $\bar s$
antiquark is in its ground state. The configurations of the
$uuds\bar s$ system, which can agree with the empirical
indications that $\mu_s$ is positive, and that $\Delta_s$ is
negative and smaller in magnitude than $\mu_s$ are the
configurations with $[31]_X[22]_S$ and $[31]_X[31]_S$ that have
$J=1$. These configurations do not include the conventional
$K\Lambda^0$ and $K\Sigma^0$ configurations, which yield negative
signs for $\mu_s$ \cite{musolf,forkel,hannelius,chenxs}.

The configuration $[31]_X[4]_{FS}[22]_F[22]_S$ stands out by the
fact that its energy is some 140-200 MeV lower than any other in
the flavor-spin interaction model of Eq.(\ref{flavorspin}).
Moreover its energy is more than 240 MeV lower than the lowest
energy configuration with the $\bar s$ in the $P-$state, which
would correspond to the $K\Lambda^0$ loop fluctuations. The lower
excitation energy should increase the probability of this
configuration as a component in the proton. Two recent diquark
cluster models \cite{jaffe,zahed} proposed for the tentative
$\theta^+$-pentaquark correspond to similar $uuds$ configurations as
$[31]_X[4]_{FS}[22]_F[22]_S$ for the $uuds\bar s$ component in the
proton with the common feature that the $\bar s$ quark in its ground
state, and hence similar values for the strangeness spin and
magnetic moment for the proton \cite{zou}.

Another interesting point worth noting is that a negative value
for the strange electric form factor $G^s_E$ is hinted at by the
data \cite{G0}. This indicates that the strange quarks have a
larger rms radius than the anti-strange quarks in the proton. The
$uuds\bar s$ configurations that give the empirical signs for the
strangeness spin and magnetic moment do have that feature, as the
strange antiquark is in the ground the $uuds$ component is in the
orbitally excited state.

The possible smaller than expected role of the long range
$K\Lambda^0$ and $K\Sigma^0$ fluctuations, which lead to negative
values for $\mu_s$, has recently been analyzed in
Ref.\cite{donoghue}. The conclusion drawn in that work is that the
main contribution to $\mu_s$ from these ``chiral loops'' arise
from their high momentum components, the determination of which a
priori falls outside of the convergence range of the effective
chiral field theory approach. As a consequence the loop expansion
as calculated by effective field theory methods may not give
realistic results. Recent numerical QCD lattice results that have
been obtained in the quenched approximation have given both
positive \cite{lewis} and negative results for $\mu_s$
\cite{leinweber} and therefore have not settled the issue of the
sign of $\mu_s$.

The diquark cluster configuration or similar
$[31]_X[4]_{FS}[22]_F[22]_S$ $uuds$ configurations for the
$uuds\bar s$ system all lead to positive values for the strangeness
magnetic moment, negative values for the strangeness spin or axial
form factor and negative strangeness electric form factor of the
proton at low values of momentum transfer. Moreover they also give
a natural explanation for the mass ordering of the
$N^*(1440)1/2^+$, $N^*(1535)1/2^-$ and $\Lambda^*(1405)1/2^-$
resonances by
admixture of large pentaquark components \cite{helminen}, which is
otherwise very puzzling in the conventional $3q$ constituent quark
model. In the diquark cluster pentaquark configuration
\cite{jaffe,zhusl}, the $N^*(1440)1/2^+$ is composed of the
$[ud][ud]\bar d$ with two $[ud]$ scalar diquarks in relative
P-wave, the $N^*(1535)1/2^-$ is composed of the $[ud][us]\bar s$
with diquarks in the ground state, and $\Lambda^*(1405)1/2^-$ is
composed of the $[ud][us]\bar u$ and $[ud][ds]\bar d$. The large
admixture of $[ud][us]\bar s$ in the $N^*(1535)1/2^-$
resonance makes it
heavier than the $N^*(1440)1/2^+$ and the $\Lambda^*(1405)1/2^-$
\cite{helminen} along with a large coupling to the channels
involving strangeness, such as $N\eta$ and $K\Lambda$
\cite{liubc}.

All facts and discussions given above suggest that there are
significant $qqqq\bar q$ components in baryons and they may be
mainly in colored quark cluster configurations rather than in
``meson cloud'' configurations.

\begin{acknowledgments}

We thank K.T.Chao, B.Q.Ma and S.Pate for the suggestion to
consider the strangeness spin of the proton and useful
discussions. C. S. An thanks S. Zhou for helpful
discussions.
B.S.Zou acknowledges the hospitality of the Helsinki
Institute of Physics during course of this work. Research
supported in part by the Academy of Finland grant number 54038 and
the National Natural Science Foundation of China under grants Nos.
10225525 \& 10435080.

\end{acknowledgments}

\end{document}